%% file: main.tex
\documentclass[sigconf,authorversion,nonacm]{acmart}

\AtBeginDocument{%
  \providecommand\BibTeX{{%
    \normalfont B\kern-0.5em{\scshape i\kern-0.25em b}\kern-0.8em\TeX}}}

\renewcommand{\paragraph}[1]{\vspace{.6em}\noindent\textbf{#1}\hspace*{.5em}}
\newcommand{\system}[0]{\textsc{Co-Pilot}\xspace}

\usepackage{enumitem}
\usepackage{xcolor}
\usepackage{multirow}
\usepackage{array}
\usepackage{siunitx}
\usepackage{graphicx}

\renewcommand{\paragraph}[1]{\vspace{0.2em}\noindent\textbf{\textit{#1}}\hspace*{.3em}}

\newcommand{\specialcellbold}[2][b]{%
  \bfseries
  \sisetup{text-rm=\bfseries}%
  \begin{tabular}[#1]{@{}c@{}}#2\end{tabular}%
}

\usepackage{xspace}

\usepackage{balance}

\begin{document}

\title[A Piece of Theatre]{A Piece of Theatre: Investigating How Teachers Design\\LLM Chatbots to Assist Adolescent Cyberbullying Education}

\settopmatter{authorsperrow=3}

\author{Michael A. Hedderich}
\affiliation{%
  \institution{Cornell University}
  \country{USA}
  }
\email{mah499@cornell.edu}

\author{Natalie N. Bazarova}
\affiliation{%
  \institution{Cornell University}
  \country{USA}
  }
\email{bazarova@cornell.edu}

\author{Wenting Zou}
\affiliation{%
  \institution{The Pennsylvania State University}
  \country{USA}
  }
\email{wpz5135@psu.edu}

\author{Ryun Shim}
\affiliation{%
  \institution{Cornell University}
  \country{USA}
  }
\email{rs2279@cornell.edu}

\author{Xinda Ma}
\affiliation{%
  \institution{Cornell University}
  \country{USA}
  }
\email{xm238@cornell.edu}

\author{Qian Yang}
\affiliation{%
  \institution{Cornell University}
  \country{USA}
  }
\email{qianyang@cornell.edu}

\renewcommand{\shortauthors}{Hedderich et al.}

\begin{abstract}
\input{paper_abstract}

\end{abstract}



\begin{teaserfigure}
  \vspace{-0.3cm}
  \centering
   \includegraphics[height=6.3cm]{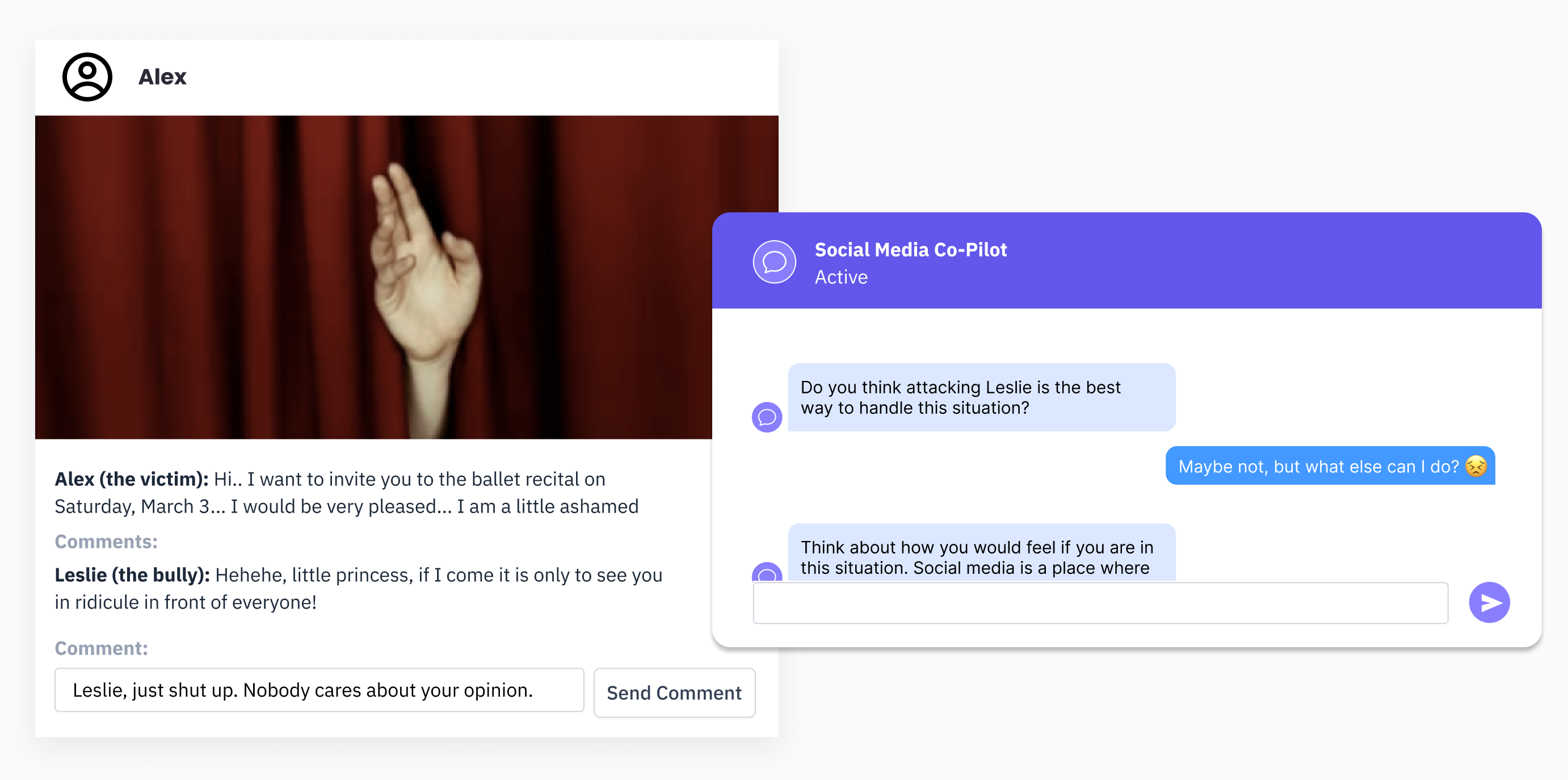}
   \captionsetup{width=.9\linewidth}
   \caption{Our prototyping platform for students learning upstanding against cyberbullying on social media. The educator can build a chatbot based on LLM-Chains that converses with the student about their bystander actions. We utilize this system as a probe to understand what levers teachers need to build chatbots that are helpful teaching tools for adolescent cyberbullying education.}
   \label{fig:platformUI}
   \Description{A figure showing our prototyping platform for students learning upstanding against cyberbullying on social media. On the left side, a social media post is visible by a user named Alex (marked as victim) who invites people to their ballet recital. Leslie (marked as bully) left a comment below the post making fun of Alex. The student bystander (user) commented "Leslie, shut up. Nobody cares about your opinion." On the right side, a chat window is visible titled "Social Media Co-Pilot". The chatbot asked the student "Do you think attacking Leslie is the best way to handle this situation?" The student response is "Maybe not, but what else can I do? [worried emoji]". One can see that the conversation continues after that.
}
   \vspace{0.2cm}
\end{teaserfigure}

\maketitle

\input{paper_body}

\balance{}

\bibliographystyle{ACM-Reference-Format}
\bibliography{ref/ref,ref/cyberbullying,ref/convomodeling, ref/authorPriorWork,ref/botdesigner}

\clearpage
\appendix
\input{paper_appendix}

\end{document}

%% file: paper_abstract.tex
Cyberbullying harms teenagers' mental health, and teaching them upstanding intervention is crucial. Wizard-of-Oz studies show chatbots can scale up personalized and interactive cyberbullying education, but implementing such chatbots is a challenging and delicate task. We created a no-code chatbot design tool for K-12 teachers. Using large language models and prompt chaining, our tool allows teachers to prototype bespoke dialogue flows and chatbot utterances. In offering this tool, we explore teachers' distinctive needs when designing chatbots to assist their teaching, and how chatbot design tools might better support them. Our findings reveal that teachers welcome the tool enthusiastically. Moreover, they see themselves as playwrights guiding both the students' and the chatbot's behaviors, while allowing for some improvisation. Their goal is to enable students to \textit{rehearse} both desirable and undesirable reactions to cyberbullying in a safe environment. We discuss the design opportunities LLM-Chains offer for empowering teachers and the research opportunities this work opens up.

%% file: paper_body.tex
\newpage

\section{Introduction}

Many adolescents have experienced cyberbullying, such as offensive name-calling, purposeful embarrassment, physical threats, and sexual harassment~\cite{vogels22TeensCyberbullying,irwin2020SchoolCrime}. Instances of cyberbullying are associated with youth depression, self-harm, and even suicide attempts ~\cite{kiriukhina2019cyberbullying,kowalski2011cyber,machmutow2012peer,price2010cyberbullying,schneider2012cyberbullying}. Large language models (LLMs) pose the risk of increasing the level of toxic online interactions even more~\cite{weidinger22TaxonomyRisks}, further jeopardizing youth's online safety and digital well-being. 
The intervention of bystanders, so-called upstanding, is an effective approach to support the victims~\cite{dominguez2018systematic,xiao2022AdolescentsNeeds}, but adolescents struggle in taking this role~\cite{allison2016cyber,zych2019children,desmet2012mobilizing}. It is, therefore, an important skill to learn and practice for digital interactions. Faced with a wide teacher shortage~\cite{dee2017teacherShortage}, especially in subjects that teach upstanding to cyberbullying like technology or health class~\cite{TeacherShortageAreas}, it is doubtful that students can receive enough personal attention to learn how to be upstanders.  

Teacher-built chatbots could scale up personalized instruction about how to upstand to cyberbullying~\cite{ueda2021cyberbullying,gabrielli2020chatbot,piccolo2021chatbots,cohen2018education,milosevic2023effectiveness}. While promising, previous research findings were limited to primarily Wizard-of-Oz studies. Translating them into actual chatbots that have an impact in the classroom requires solving technical issues around lack of data~\cite{sketchingNLP-chi19,kolchenko2018can} and necessitates that the chatbot fits into the wider curriculum~\cite{ghamrawi2023exploring,kolchenko2018can,kupperstein23AICantReplace}. Giving teachers control of LLM-based chatbots could solve both.

LLM-Chains give non-AI-experts the ability to build LLM applications with fine-grained control, but it is unknown if and how they can address the teachers' needs. LLMs drastically reduce training data requirements and with LLM-Chains, non-AI-experts can design a flow of individually configured LLMs to solve a larger task~\cite{prompt-chainer}. It is thus a promising approach for teacher-built chatbots. For chatbots, LLM-Chains have, however, only been evaluated on simple toy tasks so far and it is unclear if they can enable teachers to build complex chatbots that teach teens upstanding skills. 

In this work, we investigate to what extend LLM-Chains are a suitable approach to empower teachers to build chatbots that fit into their upstanding-to-cyberbullying education and what other kinds of support (or "levers") they need. We have developed a prototyping platform to evaluate conversational AI interventions that cultivate teen upstanding behaviors (Figure~\ref{fig:platformUI}). Leveraging this platform, we built a system as a probe and invited 13 middle school teachers to explore building a chatbot, collecting their experiences through think-aloud and interviews, which allowed us to gain their in-depth perspectives. With our probe, the teachers could gain hands-on experience building and interacting with the chatbot, thus providing deeper insights into their needs than discussing purely hypothetical situations.

Our findings show that teachers' needs for levers reflect their larger chatbot design goal: \textit{To construct a piece of educational theatre, where teens learn by rehearsing different upstanding behaviors in the social situation surrounding concrete instances of cyberbullying.} Teachers perceive their role as "playwrights" wanting to write a script for role-play social situations, ensuring that the chatbot guides students to specific behaviors while allowing students to explore different perspectives. This mindset shapes their needs for levers to further personalized instruction. To give just one example, LLM-Chains enable teachers to customize the chatbot to their class. However, new levers are necessary to allow for more controlled improvisations so students can practice upstanding more concretely, applying their knowledge to commonly encountered situations. We discuss the implications of these findings for designing levers that enhance the instructional value of chatbots for cyberbullying interventions and identify new research questions that still need to be answered in the context of chatbot use for classroom instruction.

This paper makes two contributions. First, it presents a rare description of how teachers envision using chatbots in their classrooms for K-12 prosocial online behavior education and furthers our understanding of what design and technical components can help them reach their goals. Second, it identifies new research and design opportunities about how LLMs and chatbot design tools can deliver on teachers' needs and ensure that chatbots can have an actual impact in the classroom. While LLMs are often seen as disruptive to teachers' educational and evaluative work~\cite{birenbaum2023chatbots,lo2023impact}, our work offers a complimentary perspective on how LLMs can augment it by delivering teacher-orchestrated and student-improvised personalized instruction.

\section{Related Work}

This section discusses the importance and difficulties of teaching about cyberbullying, as well as the current state of teacher-designed chatbots for this purpose.

\subsection{Teaching Adolescents about Cyberbullying and Bystander Intervention}

Cyberbullying is a form of online aggression intentionally and repeatedly carried out against victims who are unable to defend themselves~\cite{van2017thinking_Royen}. In contrast to offline bullying, cyberbullying can exhibit more complex social dynamics~\cite{Law12ChangingFace} and incorporate, as part of their attacks, a rich array of media, such as texts, photos and videos~\cite{li2007bullying}, and include manipulated imagery and deepfakes~\cite{turan2021deepfake, busacca2023deepfake}. Because the power imbalance is at its heart, cyberbullying is known to further existing social inequalities and deplete the mental health of children and adolescents, especially those from minority groups~\cite{kiriukhina2019cyberbullying,kowalski2011cyber,machmutow2012peer,price2010cyberbullying,schneider2012cyberbullying}. Addressing the needs of the adolescent victims goes beyond content moderation on social media platforms and requires a consideration of emotional impacts, victimization, and the involvement of social circles~\cite{xiao2022AdolescentsNeeds}.

Bystander intervention is widely recognized as a crucial antidote to cyberbullying and its disastrous effects on youth (see review \cite{dominguez2018systematic}).
Many U.S. students experience bullying online~\cite{irwin2020SchoolCrime}, but only a small minority tell an adult or a school teacher~\cite{patchin2012cyberbullying}. 
In this context, whether bystanders choose to reinforce a bully, stay silent on the sidelines, or support the victim becomes especially important.
Bystander actions can be public or private, subtle or direct, ranging from flagging the problematic comment to publicly defending the victim or confronting the bully~\cite{salmivalli1996bullying,difranzo2018upstanding}.

To understand the problem of bystander inaction, researchers have conducted surveys \cite{patchin2012cyberbullying} and qualitative studies such as interviews, focus groups, and controlled experiments~\cite{desmet2014determinants,desmet2012mobilizing}. Most studies have drawn on Darley and Latane's \textit{Five Stages of Bystander Intervention} framework~\cite{darley1968bystander,latane1970unresponsive}. According to this framework bystanders must first 1) notice the event, 2) appraise it as an emergency, 3) accept responsibility, 4) have the knowledge and skills on how to intervene, and 5) act. A related theoretical approach -- the situational-cognitive model of bystander behavior~\cite{casey2017situational} -- extends the bystander intervention model by accounting for additional cognitive influences (e.g., attitudes toward intervening and perceived norms for intervening), group affiliation factors, and target/perpetrator factors. These additional factors capture the influence of the social environment, which poses many perceived barriers to intervening in the eyes of adolescent bystanders.

Indeed, previous research has shown that adolescent bystanders face challenges at almost every step leading to the bystander intervention action ~\cite{allison2016cyber,zych2019children}. 
For example, they do not always appraise bullying as an emergency because the consequences of the incident for the victim, the offender, and other witnesses are often not instantly visible~\cite{barlinska2013cyberbullying,bastiaensens2015can}.
Adolescent bystanders receive little encouragement from their social environment to be upstanders~\cite{dominguez2018systematic,olenik2017bystanders}. Moreover, strong evidence indicates that their actions are highly dependent on contextual factors, such as social cues from peers and adult figures, that they are expected to act prosocially~\cite{desmet2014determinants,desmet2012mobilizing}. In contrast to offline bullying, specific aspects of online interactions, such as its asynchronous nature and large community sizes, might further inhibit upstanding behavior~\cite{allison2016cyber}. Finally, youth often lack the skills to execute bystander intervention strategies in practice ~\cite{desmet2012mobilizing}. 

Considering the need for intervention and the difficulty the youth face in performing it, it is crucial that adolescents learn strategies for upstanding. \citet{midgett2018rethinking_STAC}, e.g., created STAC, an educational program that teaches middle schoolers to develop knowledge of specific strategies to act as peer advocates. For example:

\begin{itemize}[leftmargin=*,itemsep=0cm]
\item “\textit{Accompany others}”: Reaching out to and supporting students who were the target of bullying;
\item “\textit{Coaching compassion}”: Gently confronting the bully to foster empathy toward the victim and communicating that the bullying behavior is unacceptable.
\end{itemize}

These speech acts exemplify how conversations can simultaneously provide knowledge and social guidance, thereby effectively improving bystander skills and behaviors. Further, by guiding the youth bystander through these steps, teachers could help the youth bystander practice multiple upstanding skills as the conversation unfolds. What strategy to use, however, depends on the student, and training activities are instrumental in helping students learn and practice appropriate strategies ~\cite{midgett2018rethinking_STAC}.

\subsection{Teachers Creating Chatbots for Teaching}

To scale up successful conversational guidance like STAC, chatbots could become impactful educational tools. Conversational AI technology has the potential to provide personalized and empathetic guidance to adolescents, helping them become more effective prosocial bystanders. Just as one bystander’s response to cyberbullying could empower others and help curb online aggression~\cite{Aleksandric-bystander-first-response-www23,bastiaensens2015can,allison2016cyber}, a thoughtfully designed conversational AI system likewise has the potential to mobilize young people to intervene safely and effectively. 

Researchers have started creating proof-of-concept chatbots that teach youth bystander intervention strategies~\cite{ueda2021cyberbullying,gabrielli2020chatbot,piccolo2021chatbots,cohen2018education,milosevic2023effectiveness}. These works, largely based on Wizard-of-Oz, have repeatedly shown that chatbots have the potential to guide youth bystanders to action, although none of the proposed chatbots have been implemented or evaluated with real users after a period of use. Despite its promises, bringing such conversation AI agents to the classroom still faces both conceptual and technical barriers.  

To achieve an impact in schools, chatbots need to fit into the larger curriculum and become part of the educational process. Researchers have been advocating for the inclusion of teachers in the design process of learning tools~\cite{moral2021approach}. A chatbot alone cannot replace a teacher, rather, it can enhance their teaching practice and should be seen as a new tool that supports teachers~\cite{ghamrawi2023exploring,kolchenko2018can,kupperstein23AICantReplace}. Furthermore, involving teachers in the design process has the potential to elevate their adoption of new technologies~\cite{durall2020co}. Thus, it is crucial that the viewpoint of the teacher is considered in the design and adoption process and that teachers are given control over the chatbots. The individual teacher needs to be able to adapt the chatbot so that it fits into their curriculum and becomes a useful aid to them.

Building a chatbot to help youth upstand to cyberbullying is also challenging from an AI perspective. Adolescent cyberbullying is often characterized by relational aggression (e.g., ``\textit{You are not one of us!}'') rather than explicit language~\cite{pronk2010relational_agreesion,wijesiriwardene2020alone}, making it harder to build AI to detect, much less respond to it appropriately.
Moreover, the AI needs to be empathetic, engaging, and responsive to the teen's behaviors. It also needs to monitor and regulate the escalation of emotions, considering the sensitive nature of a conversation about cyberbullying. Furthermore, lack of data, limited ML performance, and canned responses have been a longstanding issue for chatbot interfaces~\cite{sketchingNLP-chi19,kolchenko2018can}, and this is likely also limiting the advancement in chatbots for youth bystander intervention.

\subsection{Creating Controllable LLM Chatbots}

Teacher-built chatbots based on large language models could address both of the aforementioned issues, providing better chatbots from a technical perspective while ensuring that the chatbot fits into the classroom. 

LLMs have revolutionized the field of Natural Language Processing (NLP) and could help overcome the aforementioned technical chatbot challenges. LLMs can better generalize to new domains requiring only a small set of instructions and examples of desired interactions, so-called prompts~\cite{GPT3-LMs-are-FSL}. Prompting LLMs thus offers an exciting new approach to chatbot development, shifting the focus from a data question to a design question.

While prompted LLMs advance the field of chatbot design, they also bring new challenges. A core issue is controlling the chatbot's behavior, where prompting seems even less reliable than the previous ML-based design approaches~\cite{liu2021pretrain}. While guidelines for designing effective prompting exist~\cite{openAI-prompting-best-practice,promptsource}, understanding how prompts impact the output of LLMs remains an open research area in NLP~\cite{T0-Sanh2021,liu2021pretrain}. Particularly, non-AI-experts struggle when designing chatbots, suffering from both the fickleness of the prompting mechanisms~\cite{herdingAIcats} and misunderstanding the prompting capabilities, such as overgeneralizing from a single example~\cite{zamfirescu2023johnny}. 

LLM-Chains can make LLM-based chatbots more controllable but they need further evaluation. By chaining independently prompted LLM components together, the users feel more in control of the system~\cite{AIchain-CHI22}. With PromptChainer~\cite{prompt-chainer} non-AI-experts can visually design LLM-Chains, connecting LLM components in a structured flow and specifying the functionality of each component with examples. Participants in the PromptChainer study successfully built such chains, including those for a chatbot. This promising evidence suggests the utility of this approach for giving teachers control over LLM-based chatbots. However, the previous study only considered a simple music chatbot that processed one step of user interaction. What is currently missing is an evaluation of complex conversations, as one would expect from a dialogue about cyberbullying. 

The advancements in LLMs might make educational chatbots that help youth learn and practice upstanding skills a reality from a technical perspective, and LLM-Chains could potentially give teachers control over the chatbots so that they could use them in a way that fits their individual teaching and curriculum needs. This raises the question of how they want to utilize and control the chatbots for teaching about cyberbullying, how far LLM-Chains can already fulfill these requirements and what additional levers teachers need to make chatbots effective tools in their classroom. Answering these questions is our aim in this work.

\section{Method}

The goal of this study is to understand how teachers want to use chatbots for teaching youth to upstand to cyberbullying and to identify what technical and design levers they need to accomplish this task. Our aim is to guide the future development of chatbot tools to ensure that they can become implementable in the classroom.

With this goal in mind, we developed a chatbot building and testing tool for educational social media settings, which we call \system. We use this tool as a design probe~\cite{boehner2007probes} and conducted a user
study incorporating components of think-aloud, contextual inquiry and interviews. We chose this approach as our goal was to deeply understand the teachers' needs for instructional chatbot design, usage, and implementation, as well as to uncover new opportunities through teachers' perspectives. Given that LLM-based chatbots are a recent technique and chatbots in general are a novel tool in education, few teachers have experience using them. Therefore, we opted for a probe so that the teachers can gain hands-on experience building and interacting with the chatbot. We decided to let the teacher build their own chatbot from scratch as this gives the teacher a better understanding of how the chatbot works, removing some of the blackbox character of AI systems. Providing the teachers with more experience with and understanding of chatbots helps us gain deeper insights than conducting interviews about only hypothetical scenarios. Our collected data is a combination of observations of participants' interactions with the probe, their self-reported views, as well as opinions elicited through interview questions.

We will now give details on the probe (Section \ref{sec:probe}), the user study (\ref{sec:userstudy}) and the data analysis process (\ref{sec:dataanalysis}).

\subsection{Designing a Chatbot Building and Testing Tool as a Probe}
\label{sec:probe}

This subsection presents the design and implementation of the probe, which consists of a chatbot builder and a chatbot tester.

\paragraph{Design goals.~} Three goals are at the foundation of our probe:

\begin{enumerate}
    \item Without prior experience, the teacher should gain an understanding of how the chatbot system works and be able to shape the chatbot behavior.
    \item The teacher should be able to evaluate their chatbot, testing it with their own assumptions while also being confronted with external inputs.
    \item The technical burden and workload should be minimized for the teacher so that they can focus on the ideas rather than the process details. This enables us to observe more intuitive behavior and open-ended thought processes.
\end{enumerate}

With these goals in mind, we designed \system to have two core parts that teachers will use:

\begin{enumerate}
    \item Chatbot Builder: The teacher can design a chatbot without writing code or prompts. Instead, they connect graphical elements to shape the dialogue flow and provide example texts to define specifics.
    \item Chatbot Tester: The teacher can take the role of a student and interact with the built chatbot on a cyberbullying scenario on social media. The teacher is also presented with possible student answers to the chatbot from different student simulations to assist them with the testing process. The teacher can use those answers instead of their own.
\end{enumerate}

\begin{figure*}
    \includegraphics[width=0.7\textwidth]{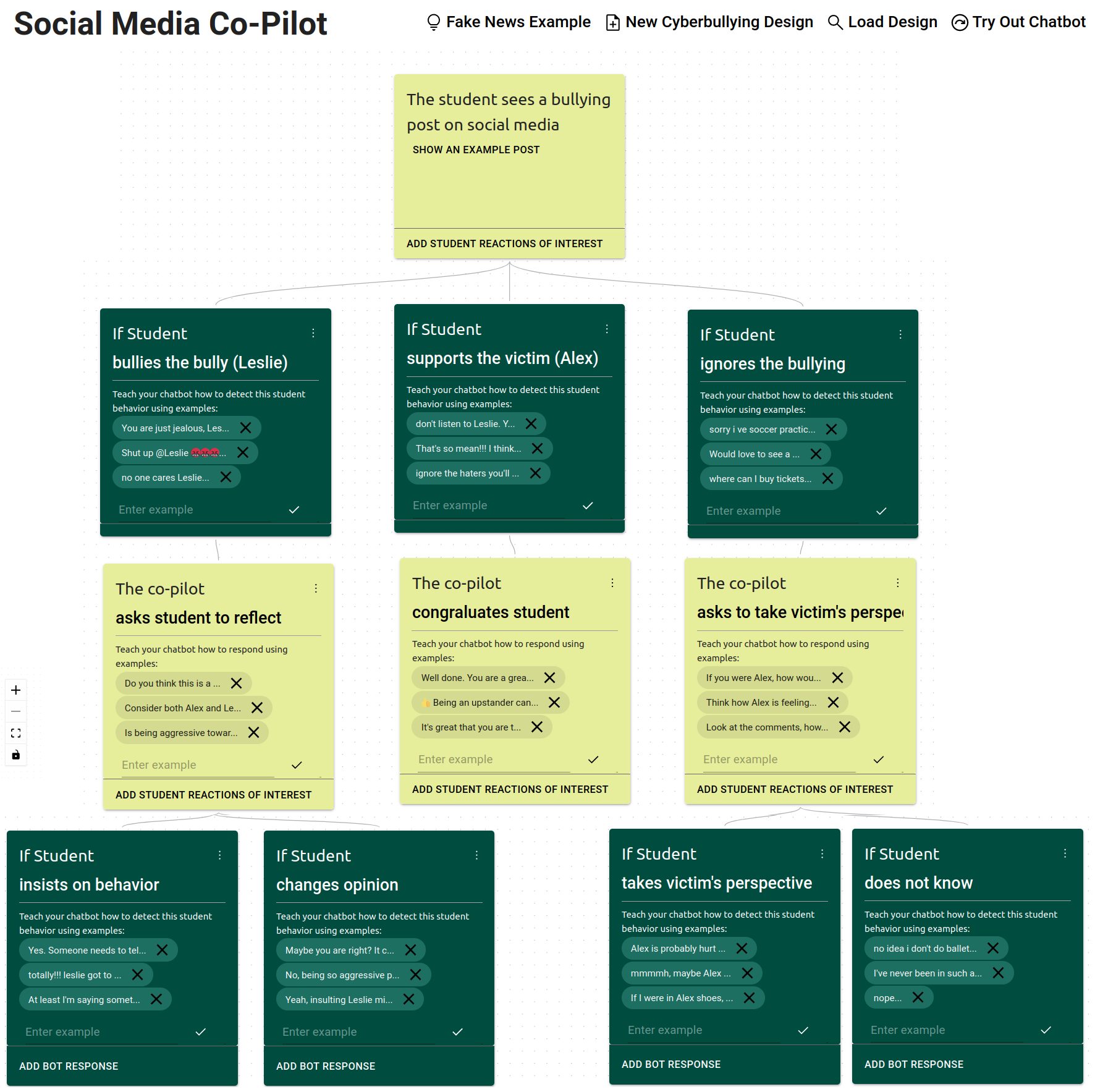}
    \caption{\system ChatbotBuilder interface showing the beginning of a dialogue flow for the cyberbullying scenario. Teachers define the possible behaviors of their students in each situation (green components) and the reaction the chatbot should give (yellow components). The teacher specifies example utterances for both types of components (chip elements). \label{fig:dialoguetree}}
    \Description{A figure showing the Co-Pilot ChatbotBuilder interface showing the beginning of a dialogue flow for the cyberbullying scenario. The structure is as defined in the paper with student behavior and chatbot reaction components building a dialogue tree. The root node lists the text "The student sees a bullying post on social media". It has three child nodes: "If student bullies the bully", "If student supports the victim", "If student ignores the bullying". Each of these nodes has a child node defining how the chatbot should reply, respectively: "ask student to reflect", "congratulates student", "asks to take victim's perspective". It is visible that the dialogue tree continues after that with components describing the possible follow-up behaviors of the student.}
\end{figure*}

\paragraph{Design of the Chatbot Builder.~} The Chatbot Builder facilitates the creation of chatbots for educational purposes, allowing the teacher to operate at two levels of abstraction~\cite{sketchingNLP-chi19}. Firstly, at the \textit{dialogue flow} level, the Chatbot Builder consists of two types of components: a) The \textit{student behavior components} where the teacher outlines the possible behaviors they expect from a student at each conversation step. b) The \textit{chatbot reaction components} where they specify how the chatbot should react to each of these behaviors. Connecting these components results in a dialogue tree, like in Figure~\ref{fig:dialoguetree}, which defines the back-and-forth chat conversation between chatbot and student. This structure allows the teacher to define controlled conversation strategies over multiple turns.

Secondly, is the \textit{utterance} level. The teachers define example texts for each of the above-introduced components. For a student behavior component, the teacher provides examples of what a student with a specific behavior (like bullying, agreeing, or questioning) might write in this particular situation. For the chatbot reaction component, the teacher crafts a set of texts that are exemplary for how they want the chatbot to answer. 

This two-level design for LLM-chains, as well as the abstraction of the prompts, are based on the PromptChainer approach by \citet{prompt-chainer}. There, predefined LLM components can be visually connected to a chain or tree structure. Their work encompasses editable LLM components which include input, transformation, output and branching/classifier components. For our setting, we adapted their approach to support multi-turn conversations where user inputs (by future students) occur multiple times. To ease the building process for teachers, we also significantly simplified their design while still being functional for our chatbot use case. We reduced the number of components from eight to the aforementioned two. We merged their input and classifier components into a single student behavior component, and our reaction component could be seen as a specialized version of their "Generic LLM." We also removed the tracking of incoming and outgoing texts across components, letting the teacher define independent examples. Last but not least, the user interface (UI) design of our components provides specific guidance on the type of input requested from the teacher.

The system uses the dialogue flow and utterances that the teacher designed and converts them into an interactive chatbot. It builds prompt-based classifiers based on the student behavior components that identify at each split point of the dialogue tree, given a student input, what path to take. The system uses the chatbot reaction components as few-shot examples for a prompt-based text generator that creates the chatbot's answer. Note that this process, including the specific prompts, is not visible to the teacher so they can concentrate on the chatbot's design.

\paragraph{Design of the Chatbot Tester.~} The Chatbot Tester gives the teacher the opportunity to test how the chatbot they have built would interact with students by playing the role of a student bystander. We use a social media scenario to guide the conversation toward the cyberbullying setting, as visualized in Figure~\ref{fig:platformUI}. The bystander is presented with a social media post featuring an exchange between a victim and a bully, and the bystander can comment on this post. The chatbot starts the conversation based on the bystander's comment on the social media post. It opens a chat window, mimicking how the bystander student might receive a personal message (a ``DM'') on a social media platform. The bystander can answer the chatbot, and the conversation between the chatbot and the bystander unfolds.

In our study, the teacher took the perspective of the student bystander to test the chatbot. They could write comments on the social media post, as well as direct answers to the chatbot. Their inputs and the chatbot's reactions allowed them to examine how their design would be reflected in the realized chatbot. It also enabled them to test the limits of the chatbot and try out new ideas, thus gaining a better understanding of its behavior and possible impact on learning.

To enable teachers to experience less strictly designed interactions, the conversation could continue even after the chatbot reached the end of the dialogue flow created by the teacher. When the system reached a leaf component in the dialogue tree, it continued to respond to further messages. The LLM-based chatbot generated new responses by taking into account the teacher's instructions defined in the last component as well as the new bystander message inputs.

Additionally, we provided student simulations as an external input to the teacher that might challenge their assumptions. These student simulations were shown to the teacher as suggested student comments and responses. The teacher could use them instead of their own texts. The text suggestions were generated by LLMs, which were prompted to represent a specific student behavior. In contrast to the chatbot the teacher built, LLM-generated suggestions did not use any controls from our side. Instead, the LLM generated a text solely based on a short behavior description and the conversation history thus far. We implemented three student behaviors, namely, a student attacking the bully, a student supporting the victim (upstander), and a student ignoring the bullying (passive bystander). Although we considered using answers created by real students during the study design, we opted not to because presenting only pre-collected student answers might not match the conversation flow designed by the teacher. Using live student responses would have also been sub-optimal because it would have moved the study's focus away from the empirical evaluation of the teacher's exploration. 

We aimed to provide a realistic-looking social media scenario in both design and content. We based the social media post and the bully's comment on the ballet scenario from \cite{Italian-WhatsApp-Data} translated into English and using gender-neutral names. Throughout the design stage, we consulted with two teenagers and integrated their feedback into the study design.

\paragraph{\system~Implementation.~} We implemented \system as a React-based web application with a Python Flask backend and relied on OpenAI's \texttt{GPT-3.5} models as LLMs. For the chatbot, we used \texttt{Text-Davinci-003}, as it mimicked the teachers' examples more closely without requiring additional prompting in pilot tests. For the student simulations, we used \texttt{GPT-3.5-Turbo} (ChatGPT) for its more adaptive answering behavior. We give further implementation details in the Supplementary Material. At the time of implementation, the more recent \texttt{GPT-4} and \texttt{LLaMA2} models were not yet available to us. However, we argue that our approach is generally independent of the latest large language model as we are interested in the teacher's needs and not the exact system performance.

\subsection{User Study Design.}
\label{sec:userstudy}
To understand teacher's needs concerning chatbots and how they want to use them as tools for teaching teenagers about cyberbullying, we invited 13 teachers to use \system
and think-aloud.

\paragraph{Participants} All recruited participants ($N=13$) had experience teaching in middle school. To avoid excluding participants based on coding or prompting experience, the probe did not require any technical experience from the participants. Our sample size was chosen in line with prior work~\cite{faulkner2003beyond,sauro2016quantifying}.

All participants except P1 had experience in teaching digital citizenship. Our participant pool thus contained teachers who were invested in teaching about bystander interventions and cyberbullying. Cyberbullying and upstanding intervention are taught as part of different subjects, like health, technology or digital citizenship. Participants had, therefore, diverse teaching backgrounds and roles. Table~\ref{tab:participants} lists these as well as the teachers' experience levels. We obtained IRB approval before starting the study. All participants received a \$25 voucher for their time.

\newcolumntype{C}[1]{>{\centering\let\newline\\\arraybackslash\hspace{0pt}}m{#1}}

\begin{table*}[t]
    \centering
    \caption{Interview Participants. All participants had experience teaching in middle school. Depending on the school, cyberbullying was covered in different subjects, like technology, health or digital citizenship. ICT facilitators were teachers who also taught other teachers about digital citizenship methods and coordinated corresponding programs.}
    \label{tab:participants}
    \begin{tabular}{C{0.05\textwidth} C{0.20\textwidth} C{0.20\textwidth} C{0.10\textwidth} C{0.15\textwidth}}
    \toprule
    {\specialcellbold{ID}} & \textbf{Role}  & \textbf{Current Class on Cyberbullying} & \textbf{Years Teaching} & \textbf{Region} \\
    \midrule
    P1 & Teacher & N/A & >30 & US\\ 
    P2 & Teacher & Info Technology & >10 & US\\ 
    P3 & Teacher & Health & >10 & Canada\\ 
    P4 & Teacher & STEM Program & >5 & US\\ 
    P5 & ICT Facilitator & Digital Citizenship & >5 & US\\ 
    P6 & ICT Facilitator & Digital Citizenship & >5 & US\\ 
    P7 & Head ICT Facilitator & Digital Citizenship & >10 & Southeast Asia\\ 
    P8 & Teacher & Computer Science & >20 & US\\ 
    P9 & Teacher & Health & >10 & US\\ 
    P10 & Teacher & Leadership Character & >10 & US\\ 
    P11 & Librarian & Technology & >30 & US\\ 
    P12 & Teacher & Digital Citizenship & >10 & US\\ 
    P13 & Head ICT Facilitator & Digital Citizenship & >5 & US\\ 
    \bottomrule
\end{tabular}
\end{table*}

\paragraph{Task} Participants (teachers) were shown a social media scenario featuring a case of cyberbullying. They were asked to create a chatbot that would engage in one-on-one interactions with students who were exposed to the cyberbullying situation as bystanders. The interaction would be triggered by the bystander's comment to the cyberbullying social media post, and the teacher's task was to create a chatbot who would initiate and carry on the conversation with the student (bystander). We asked the teachers to design the student behavior components according to how they would expect their students to behave. They were free to specify how the chatbot should react in each situation and how long the dialogue flow should be. 

After building the chatbot, we asked the participants to test it, taking the student's bystander perspective. Participants had full range in exploring how the chatbot reacts. They could input their own comments on the social media post and their own answers to the chatbot. Alternatively, they could use the suggested texts by the student simulations or a mixture of both. Participants could switch freely between the student simulations and reset the conversation at any time to the start point. Participants had the option to go back to the Chatbot Builder and modify their chatbot if they desired.

\paragraph{Interview Protocol} The interview started with the participant presenting their teaching background and how they teach their students about cyberbullying.

We then gave the participant an introduction to \system. To avoid biasing the participants with a pre-existing chatbot design on cyberbullying, the topic of the introduction was on fake news, a different digital citizenship topic. Each participant was first shown the social media scenario for fake news (i.e., the Chatbot Tester) followed by an exemplary chatbot design within the Chatbot Builder, highlighting the two levels of abstraction (dialogue flow and utterances).

We then asked the participant to build their chatbot. The interviewer showed the participant an example of a social media post for cyberbullying and suggested starting with defining possible behaviors they would expect from their students in this situation and how the chatbot should react. The participant was then given complete control of \system and asked to think out loud while building the chatbot. The interviewer further advised participants only when they asked for assistance. The advice was limited to helping with UI questions (such as how to move components on the screen) and the suggestion to use their teaching experience for designing the chatbot.

Once the participant indicated that they had finished building the chatbot (or after 45 minutes had expired since the start), the interviewer suggested switching to the Chatbot Tester. Again, the participant had the freedom to explore and was asked to comment on their testing. The testing continued until the participant indicated that they had finished (or after the 60-minute interview mark was reached). 

We informed the participants that the session's goal was to understand how to teach teenagers about upstanding to cyberbullying and if or how chatbots could potentially play a role there. We clarified that the probe was an early prototype and we emphasized our interest in receiving their honest opinions. During the session, we observed how the participants used and explored the \system and recorded their comments. When the participants mentioned aspects relevant to the research question during their thinking-aloud process, the interviewer asked them to elaborate. These elaborations constituted the most significant part of the collected interview data. After the exploration phase with the \system, the interviewer asked the participant a set of questions if these had not been addressed by the participant already. Specifically, we asked i) if or how they would use a chatbot in their class when teaching about cyberbullying, ii) if or how they would like to build or customize a chatbot for cyberbullying, and iii) if they could wish for new functionality or support, what would that be.

We performed the user study remotely over Zoom. The \system was hosted on a server so participants could access it on their browser during the interview.  For two participants whose schools' firewall blocked access to our probe website, the interviewer shared their screen, and the participant gave them instructions on what to do during the building and testing of the chatbot.

\subsection{Data Analysis}
\label{sec:dataanalysis}

We recorded and transcribed the user study. For each participant, two authors independently reviewed the transcript and distilled important insights from it. The union of these emergent insights was used to create affinity diagrams to synthesize and organize observations across the interviews. The inspection and labeling of affinity diagrams, which were discussed with all authors, revealed key themes and patterns. Their contents were further analyzed to categorize and prioritize the themes, as well as to merge or remove overlapping clusters. After finalizing the diagrams, two authors independently verified all findings against the original transcripts and found no discrepancies.

We chose affinity diagrams instead of grounded theory for several reasons. This method is often used in HCI and interaction design practice~\cite{lucero2015affinity,harboe15affinity}. Furthermore, our objective was not to build up a theoretical account of how teachers designed chatbots with existing tools. Instead, we followed a more practice-based approach to inform the design and application of new resources and tools by directly engaging teachers in the chatbot building and testing. The observations, combined with interview insights, revealed teachers' preferences for the design and deployment of chatbots as an instructional tool for teaching bystander intervention in the classroom. 

\section{Findings}
In line with previous work, our interviews showed the potential of chatbots in scaling up personalized and interactive teaching of bystander intervention. P11 described bystanders as individuals that \emph{"just sit and watch,"} emphasizing that many \emph{"really want to say something, but just stand there."} The introduction of chatbots challenges this passive tendency often exhibited in cyberbullying cases, urging students to take on a more proactive role.

One of the key advantages of chatbots over traditional teaching methods is the capacity to deliver immediate and individualized feedback. This quality distinguishes chatbots from conventional lessons, where several participants reported difficulties in addressing the needs of every student due to time constraints and class size. P2 praised the impact of this feature, stating \emph{"I don't think it's going to have the same effect if I wait until tomorrow to [correct] them or after I grade a paper. [The chatbot] keeps those wheels turning."} 

A teacher's task, however, is not purely instructional, with P8, e.g., describing her role ``\textit{not [as] a knowledge-giver but a moderator.}" This sentiment is reflected in the teachers' needs for the chatbots as well. Our findings from teachers building and testing chatbots with \system reveal that they did not perceive their goal as prescribing a conversation that the student would loyally carry out with the bot. Rather, we found that
\begin{enumerate}
    \item teachers wanted to design chatbots that are part of multi-participant role-plays that enable students to take on different perspectives, and
    \item by allowing the chatbot to improvise within the limits of the teacher's guidance, teachers wanted to create scenarios where students can explore and practice socio-emotional skills in a safe environment.
\end{enumerate} 

We unite these needs under the larger theme of teachers wanting to be playwrights: the teacher's role resembles a modern playwright in that they develop characters, and create role-play scenarios or plots that align with the (educational) goals. The actors (learners) are allowed to rehearse and improvise within the framework of their characters to deepen their understanding of the impact of their role's actions.

In Section \ref{finding:playwright}, we unpack the teachers' perspectives on using chatbots for teaching about bystander interventions to cyberbullying and the goals they want to achieve. Section \ref{finding:existing_lever} describes how existing LLM-Chains support these goals, while Section \ref{finding:new_levers} uncovers needs that are not yet met and what additional levers the teachers require.

\subsection{The Teacher as a Playwright}
\label{finding:playwright}
This section details the teachers' needs with regard to using chatbots for bystander intervention education.

\paragraph{Learning Socio-Emotional Skills} Teachers are not merely interested in instructing intervention steps; instead, they aspire to cultivate socio-emotional skills within their students in order to better navigate cyberbullying situations. P8 described current teaching of social media education as ``\emph{hand slapping lesson}'' just focusing on teaching students prescriptive rules. P6 identified the importance of moving beyond this form of teaching, stating that students needed to first understand the underlying issues and the harm caused by cyberbullying before teachers could address student interventions.

A more holistic approach aims to guide students in developing broader skills, such as perspective-taking and empathy, and approach intricate nuances of such situations with sensitivity. P1 offered insight into this perspective, noting that social situations involving cyberbullying are complex and multifaceted as \emph{“not 100 percent [of blame should] be placed on one person only. [...] There are at least factors from all parties that lead to this situation.”} Similarly, P11 highlighted the importance of instilling empathy amongst students, stating that \emph{“everyone today really needs to understand where the other person is coming from and have some empathy for others.”}

\paragraph{Learning Through Multi-Participant Role-Play} To help students understand the perspectives of the various stakeholders involved in cyberbullying situations, many participants suggested involving multiple chatbots and the student in a role-play scenario. The teachers saw their task in preparing these scenarios and in defining the different roles, including the bully, victim, and bystanders. The chatbots and the student would then play their roles by commenting and messaging on the social media scenario.

This role-playing approach offers a unique opportunity for students to grasp the impact of their actions in an empathetic manner. As P5 pointed out, \emph{“Usually, you just ask [students] to reflect on it and pose some questions and ask them well, how did this make someone so feel? [...] [Role-play] would be a quicker way for them to grasp the impact of their actions on someone else.”} P11 echoed this sentiment, highlighting that this approach enables students to empathize with various roles, including that of the victim, the bystander, and even the bully, stating, \emph{“This is giving someone a way of stepping in someone else's shoes in social media."}

In contrast to traditional classroom role-plays, chatbots provide a safe space for role-playing without the fear of judgment. P13 highlighted that this chatbot \emph{“allows kids to do things that they may not feel comfortable with in front of a whole group.”} Likewise, P10 pointed out that in their previous experience, students often felt compelled to clarify that their assigned role-play behavior did not necessarily reflect their real-life actions. Similarly, P11 saw the chatbot as an avenue for students to explore ``what ifs'' in a private setting.

\paragraph{Catalyzing Learning Through Repetition, Exploration \& \\ Guided Improvisation} Many participants wanted the chatbot to empower students to practice and make corrections in a safe space, providing a learning experience they could fall back on while navigating the world around them. For that, they wanted the chatbot to improvise on their instructions so that students could extensively explore challenging cyberbullying situations and try out different roles.

The chatbot gives students a platform to explore different behaviors in a safe environment. P8 and P11 acknowledged the importance of making mistakes and learning from them, mirroring the developmental stage and learning style of middle school students. Similarly, P11 recognized the impulsive nature of middle school students who are still learning how to express themselves. She saw the chatbot as an opportunity for the student to ``\emph{write inappropriate things [to] see what the chatbot responds [...] to do what they might be impulsive to do.}'' P3 even expressed a desire to encourage this and for students to experiment with different behaviors, both \emph{“confrontationally”} and \emph{“nicely,”} to observe how the chatbot responds. P9 saw the chatbot interaction also as an opportunity for the student to vent in a cathartic fashion. The teachers emphasized that the chatbot provides a safe environment for exploration, with P8 stating ``\emph{We're learning; we're supposed to make mistakes. And [students] have a safe environment here.}'' 

P6 and P8 believed that students should also encounter situations that can go awry. For instance, P6 envisioned a scenario where a student exhibits the desired upstanding behavior as taught in school, however, the bully persists. P6 elaborated stating, ``\emph{Maybe the co-pilot creates fake responses to continue the bullying [...] to help kids realize [...] sometimes it doesn't go smoothly. Sometimes you can say stop, and [bullies] don't always stop. And I think getting the kids to realize that and [...] help them realize that your first attempt may not always pan out and help them practice that.}'' This approach aims to prepare students for real-life conflicts, in which their actions may not yield straightforward or predictable outcomes.

The teachers stressed the importance of repetition within this exploration and the need for the chatbot to improvise within their guidelines to support the student's practice. P9 highlighted the value of having the chatbot reiterate statements using different phrasing. This approach is particularly beneficial because, as P9 pointed out, the students in that age group best absorb information through repeated exposure. P4 and P13 also stated that they want students to repeatedly try again, with P4 saying that they want to design the chatbot in order to \emph{“have [students] try over and over again, to recognize, what is [the students'] responsibility here”}.

The value of this approach is further underscored by P10, pointing out that compared to traditional teaching methods, the \emph{“hands-on”} role-play approach is \emph{“no longer memorization… [and is] becoming muscle memory.”} This experiential learning allows students to transform their conceptual understanding into practical, real-world applications.

\paragraph{Adapting the Chatbot} To align the chatbot with specific aspects of their school, address unique situations in their class, and match their own teaching style, teachers emphasized the importance of customizing the chatbots.

Participants wanted the chatbot to be reflective of their school and class. P5 and P9 both remarked that when they were teaching these topics, they adapted their scenarios to specific situations that happened to their students in real life to make the experience more engaging and realistic. P6 and P11 wanted to integrate references to personnel at their school so that their students could get advice tailored to them and have a more personalized experience. P8 noted the need to adapt to differing terminology between schools. P11 additionally referenced their school's foundational principles, while P7 wanted the chatbot to provide links to additional resources.

P2 noted that the chatbot's language should align with that of the students. A similar viewpoint was shared by P8, who emphasized the importance of adapting the wording to match the way students speak, considering the fast-evolving nature of their slang and its unique local forms. P9 argued that this representation of the students' language is important to increase engagement. 

Teachers emphasized that it is not merely about having a standalone chatbot; it needs to be an integral part of their teaching approach and match their personal teaching style. P9 underlined the individuality of teaching styles. They state that \emph{“Every teacher has a different style in the classroom,”} therefore, it is important to allow teachers to tailor the chatbot to align with their unique teaching styles. P9 described their own gentle approach to redirection where, e.g., P11 noted the need to send clear stop signals in certain situations, and P8 remarked that they usually added material beyond the standardized curriculum to push their students further.

\subsection{Existing Levers: LLM-Chains For Teachers as Playwrights}
\label{finding:existing_lever}

Understanding the teachers' perspective as playwrights helps to evaluate to what extent LLM-Chains can empower teachers to build chatbots that are useful teaching aids to them. We find that the LLM-Chains ability to adapt based on few examples while being controlled with the chain-structure and the flexibility of LLMs to reformulate answers are useful levers to the teachers.

\paragraph{Adapting the Chatbot} Teachers wanted to adapt the chatbot to their school, and this custom adaptation was made possible by the LLM-Chains. P11 added, e.g., a specific reference to their principal naming him in the example answers of the chatbot. This allowed the chatbot to refer to the principal during the bystander chat. 

The teachers also used the LLM-Chains to integrate their own teaching style. P6, e.g., wanted the chatbot to acknowledge positive student behavior and redirect student actions if they encouraged cyberbullying. When testing the chatbot they had built, they commented: ``\emph{I'm pretty happy with the way this chat is going, especially considering how little I put on the chatbot side.}'' P11 also expressed that it accurately conveyed what they intended to communicate and, likewise, P8 saw how the chatbot mirrored and reflected ``\emph{the same tone but in different words.}'' P8 continued stating, ``\emph{It really reflects [me]. That's really amazing. Even in those few examples -- wow.}''

P8 advocated for this level of customization, commenting on the result: \emph{“I care about the kids, and I want them to know that. [The chatbot builder] can help take what makes me special as a teacher and put it into a tool like this”}.

\paragraph{Catalyzing Learning Through Repetition} Teachers highlighted the importance of repetition when students are learning about bystander interventions to cyberbullying. The LLM-Chains allowed the teachers to define chatbots that could reformulate their example answers. The students would then be presented every time with new answers that still followed the teacher's guidance.

When testing the chatbot, teachers remarked positively about the chatbot's rephrasing. P7 stated that having ``\emph{always the same questions, the same answers [is] boring}'' and that the chatbot was useful because it answered in different ways, rephrasing the teacher's message that one should be more respectful and caring. P8 commented that the chatbot ``\emph{doesn't sound like a machine}'' and that it correctly rephrased their examples. P9 was surprised by the chatbot's ability to answer the student in repeated and rephrased form and expressed that ``\emph{every one of those responses is awesome for [the students] to hear.}'' They expanded on this point stating, \emph{“[The chatbot] is good, because every one of these responses is different [and the students are] going to read every one of those."}.

\subsection{New Levers Needed By Teachers as Playwrights}
\label{finding:new_levers}

While LLM-Chains provide some of the functionality to enable teachers to become successful playwrights, our participants also reached the limitations of this approach in several aspects, which suggests the need for new levers discussed in Section~\ref{sec:discussion}.

\paragraph{Levers That Support Playwriting} As a playwright, the teacher is tasked with narrating the behavior of students and chatbots. Among the participants in the study, there was a noticeable variation in their ability to generate examples of their behaviors. Some participants found the process of designing student behaviors and chatbot responses to be relatively easy and intuitive. P6, in particular, demonstrated a swift ability to generate responses, stating that the reason is \emph{"lots of experience working with kids and teaching, and navigating social media"}. 

However, some participants faced significant challenges. P2, P8 and P13 indicated signs of struggling when trying to verbalize examples for the student behavior components. P10 found it particularly difficult to adopt the mindset of a middle school student, stating, ``\emph{Putting yourself in the middle school age, I think makes it a little difficult because as an adult, obviously, my brain is going to work differently.}'' P12 similarly noted that they need to get back into the mind of their students. Meanwhile, P3 and P6 found it difficult to identify all possible student behaviors, with the latter stating: ``\emph{So the student joins the bullying, ignores the bullying [...] I feel like there's one more option.}''

In selected cases, the struggle of comprehensively describing the student behaviors was also reflected during the testing phase. For P3, the passive bystander behavior of one of the student simulations did not match any of the behaviors they had defined, resulting in the chatbot being unable to respond appropriately. Seeing the chatbot's reaction to the student simulation, they realized what they had missed, commenting ``\emph{Oh, why didn't I think of that?}''

The teachers know what socio-emotional skills they want to convey to their students with the chatbot, but they struggle with creating a script for the parts of middle schoolers. Many of them would benefit from supplementary support to address the challenge of accommodating students with diverse behaviors. 

An LLM could be used as a lever to provide writing support when building the chatbot. P8 requested a resource where they could pull examples from, while P2, P3 and P12 wanted suggestions automatically provided while they built the chatbot. With the right prompting, the LLM could propose student behaviors or utterances for each situation. The teacher could get inspired by these suggestions for their own writing or use them directly if they agree with them. P2 commented on the LLM output during testing that ``\emph{Somebody else [the LLM] is way more creative with our words than me.}'' This suggests that an LLM-based writing assistant could assist teachers with the script-writing process.

Besides collaborating with an AI, teachers also want to work together with their colleagues. Cooperation among teachers in the context of curricula is familiar to them, as highlighted by P5. In their school, a common planning time exists to plan lessons together, distribute tasks, obtain feedback, and share results. They expressed the desire for a similar collaboration in chatbot design. P2 also wanted to collaboratively develop the chatbot with fellow teachers, while P6, P8 and P12 emphasized the sharing of chatbots with other teachers.

Such collaboration is not limited to only teachers but could also involve students. P11 stated that the students already contributed to the teaching process by sharing their own cyberbullying experiences, and P10 emphasized that this allows them ``\emph{keeping a pulse of what's going on in our school.}'' P8 argued that the students' input is especially valuable as social media is not P8's world. They all, along with P7 and P12, wanted to leverage students' experience and insights by involving their students as feedback-givers or co-writers of the chatbot.

New levers that support the teacher in playwrighting could thus be either of technical nature, benefiting from LLM suggestions, or transfer collaborative structures already existing at schools into the chatbot-building process.

\paragraph{Levers to Guide Chatbot Improvisation} The teachers also wanted the chatbot to improvise so students could explore different behaviors in-depth. Rather than strictly adhering to scripted responses, the LLM-Chains could create a chatbot guided by the examples provided by teachers while having a degree of improvisation built-in in its interactions.

Several teachers commented positively on the chatbot taking these liberties. P6 stated, ``\emph{They're good responses. Especially because there are so many answers the student could give  [...] I think it's good that the chatbot is able to take over and recognize the different
responses and continue having that discussion [...] without me needing to pre-program everything into it.}'' P7 and P9 were surprised by the depth of the chatbot's follow-ups.

Some participants also encountered, however, limitations in the chatbot's ability to improvise. If the student continued the conversation beyond the last component defined by the teacher in the flow of the LLM-Chain, our probe proceeded to use the last teacher's instruction as guidance. For P8 and P9, this process resulted in the chatbot ending up in a conversational loop, always rephrasing the same type of answer. The teachers asked for an option to define when the chatbot should switch to a new conversational topic in such a situation. They suggested that the switch should occur once the student shows understanding of the chatbot's message or after a predefined number of repetitions.

P6 also emphasized the importance of the chatbot adhering to the predefined guidelines, expressing concerns that the chatbot might deviate too much from the intended educational path: ``\emph{I would worry that the chatbot started agreeing with the [bullying] student [...] or started veering in the wrong direction and [I would] just make sure that it stays positive.}''

While LLM-Chains are a lever that gives teachers control over the chatbot, the guidance the teachers provide is bound to the dialogue flow structure. The chatbot can improvise within this structure but struggles to go beyond it. The LLM-Chains can give the teacher fine-grained controls, but new levers are needed so that teachers can better guide the improvisation more abstractly. These new levers should allow teachers to define higher-level chatbot behaviors, such as when to move to a new conversational topic. At the same time, these new levers still need to let teachers enforce their guidelines, ensuring that the playwright stays in control.

\paragraph{Levers That Enable Multi-Participant Role-Play} Furthermore, teachers want to design role-plays with multiple participants. Supporting such interactions adds a new dimension to the chatbot design. Chatbot interactions are usually 1:1 conversations between a user and a chatbot. However, teachers were interested in having their students explore social situations that simulate interactions of multiple participants, including the victim, the bully, and other bystanders. This requires multiple chatbot participants interacting with each other and the student.

While teachers could use separate LLM-Chains to build different conversation participants, the chatbots must be aware of the other participants, their roles in the social environment, and their actions. This will require connecting the chatbots and updating their information about each other and the student while the conversation progresses. New technical levers are needed to support such interactions.

Multi-participant role-plays are also a design challenge. In our probe, teachers only needed to conceptualize the possible actions of a student and how their chatbot should react to each of them. Even then, P12 explained how they preferred to map out such branching systems first on paper. Adding multiple active roles to the scenario would require the teacher to define how each chatbot should react to the other chatbots and the possible student behaviors. Some roles might also change their behavior over time (e.g., a passive bystander becoming an upstander) and might, therefore, also adapt their interactions with the other participants. Building chatbots adept at navigating an increasingly spiraling complexity of multi-role conversations would burden the playwriting teacher. Therefore, new design levers are necessary that will enable teachers to guide chatbots in such multi-participant role-plays.

\section{Discussion}
~\label{sec:discussion}

In this section, we will first discuss our findings on teaching bystander interventions to cyberbullying through role-playing with chatbots. While teachers see their role in this context as playwrights, our findings showed that new levers are necessary to enable teachers to succeed in this role. In the following subsections, we will discuss the design and system opportunities ensuing from these findings, as well as outline future research directions to address existing research gaps in the instructional use of chatbots for teaching prosocial behaviors to adolescents.

\subsection{Teaching Prosocial Behaviors With Chatbots}

In line with previous research, our findings show that teachers want to provide personalized ways to teach bystander intervention and that chatbots have the potential to provide such teaching at scale. We also show, however, that the teachers want to go beyond providing an interactive way to learn about conversational guidance like STAC. Instead, they want to build chatbot-based role-play scenarios where students can actively explore bystander behavior.

While chatbots have been previously explored as effective instructional tools for enabling role-playing for situated, authentic, and safe learning in dialogic-centric settings~\cite{othlinghaus2020technical}, our findings provide unique insights into teachers' role as playwrights in a role-playing learning process. When teachers are playwrights, chatbots can be effective classroom aids and resources, assisting teachers in training students in prosocial behaviors necessary for upstanding against cyberbullying and confronting other digital risks. The teachers in our study, by and large, embraced the role of playwrights, viewing student-chatbot role-play as an effective tool for students to learn and practice perspective-taking, empathy, and nuanced consideration of their own and others’ actions necessary for bystander interventions to cyberbullying. What emerged from our findings is the collaborative role-playing orchestrated by the teachers but leaving room for student improvisation and experimentation in a safe conversational space. Through conversational planning and regulation, a teacher can create scripts that allow students to practice upstanding behaviors and other prosocial communication strategies in a realistic conversational exchange. Furthermore, the playwright role allows teachers to customize the learning process and learning outcomes to satisfy current and emergent student needs and connect role-playing to the curriculum goals and the rest of the school experience. 

Instead of structuring the student training mechanistically by giving students "recipes" for how to act as an upstander, the teachers emphasized the importance of developing contextual and social awareness so youth can read a cyberbullying situation in a contextually-sensitive way and respond with appropriate communication strategies. Their guidance went beyond the prescriptive chain of actions outlined in the bystander intervention model ~\cite{darley1968bystander,latane1970unresponsive} (i.e., notice an emergency, recognize it as such, take responsibility, know how to intervene, and act). Instead, teachers used scripts as opportunities to help youth develop communication and socio-emotional skills, such as social awareness~\cite{rimm2020sel}, which can be seen as overarching competencies instrumental for each stage of the bystander intervention process. In this respect, the approach taken by the teachers in our study was more consistent with the situational-cognitive model of adolescent bystander behavior~\cite{casey2017situational}, which emphasizes the embeddedness of a cyberbullying episode within social and peer contexts, and the entanglement of bystanders' actions with interpersonal relationships, social group affinities, status hierarchy, and community climate. As a result of these entanglements, bystanders experience high uncertainty about which options are socially appropriate and safe and have to contend with possible fallout from intervening. To overcome this uncertainty, bystander theorists recommend "the need for skill practice across a range of scenarios, using a variety of possible bystander responses"~\cite[p. 18]{casey2017situational}. Chatbot roleplaying enables this multifaceted practice recommended by theorists, and the teachers' scriptwriting approach guided by their practical experience working with adolescents was well-aligned with this recommendation. 

Below, we discuss the opportunities that LLM-Chains offer to the design of teacher-built chatbots and identify crucial pedagogical and technological research gaps. 

\subsection{LLMs Supporting Teachers in Playwriting}
Although teachers viewed their role as playwrights, writing the ``script" that prompts youth interventions to cyberbullying can be difficult and might require help that LLMs could provide. We identified that writing in the style of students and anticipating their possible behaviors can be a challenge for teachers, and some of them requested additional support. For the writing style, researchers have shown that LLMs can adopt different text styles, including slang and chatty forms~\cite{reif22StyleTransfer,shu23RewriteLM}. To help teachers define various possible behaviors that reflect students' uncertainty and hesitation around bystander interventions, they could utilize LLM suggestions. \citet{hamalinen23SyntheticData} used LLMs for generating synthetic user data. An LLM system might similarly be able to generate behavioral data for student exchanges, suggesting student reactions to the teacher. The teacher could then validate these synthetic data according to their experience, quickening the chatbot creation process and filling gaps the teacher might have missed. 

One needs to be, however, keenly aware of LLMs' limitations and the biases they can introduce. Language models reflect the textual data they are trained on and thus only represent the pool of existing data. Depending on the training timepoint, it is unclear if they can keep up with rapid-moving trends of teenagers, for example, with teenage slang, pop culture shifts, and social media interactions. When considering subjective opinions, researchers have already shown that LLMs are biased towards specific ideologies~\cite{felkner23LGBTQ+Bias,motoki2023PoliticalBias} and populations~\cite{durmus23GlobalOpinions,venkit23NationalityBias}. It is thus essential to understand if LLM suggestions for teachers can support them in building chatbots with a broader student representation or if the LLM causes the opposite, biasing and narrowing their design. 

Furthermore, while LLM suggestions may reflect a broader student representation, further adaptation may be needed to reflect specific geographical, socio-cultural, developmental, and other sub-group identities of students in a particular classroom. Teachers may even consider running chatbot suggestions by student helpers to ascertain their relevance, typicality, and realism. In this case, scriptwriting would become a collaborative process, with teachers orchestrating the script, but LLM and student helpers supplying and reforming textual data, as we discuss in more detail next. 

\subsection{Collaborative Chatbot Design With Teachers And Students}

Collaboratively designing the chatbot could result in learning tools that are pedagogically more inclusive and effective. Recognizing the benefits of human-centered design, researchers have been arguing for the inclusion of learners and teachers in the design process of learning tools that are pedagogically inclusive and effective~\cite{moral2021approach,durall2020co,durall2020ownership,kolchenko2018can,kupperstein23AICantReplace}. Our findings reflected these arguments showing that teachers value the input from other teachers as well as from their students. They repeatedly voiced their wish to seek out their colleagues and students when building the chatbot. Systems that support collaborative workflows where teachers can ask for feedback or share their work could support the adoption of educational chatbots as shared tools in the classroom.

A promising solution might be to use collaborative exercises with a teacher and their students working together to create a chatbot-based role-play. This kind of collaborative storytelling has been previously used in creating stories for role-playing games in classroom spaces~\cite{jones2021chasing}. Like choose-your-own-adventure books, participants can narrate different action possibilities depending on the story characters' steps. Furthermore, students' involvement in this process can also serve as an exercise in perspective-taking, critical reflection, and engagement skills~\cite{cook2017we}. Critically, bringing in student voices and perspectives will ensure that the actions and contexts created through collaborative storytelling will accommodate the actual concerns and experiences of youth involved in the process, which is critical for fostering engagement and adoption. 

\subsection{Teachers Guiding Chatbot Improvisation}
Teachers seek chatbot improvisation while maintaining control. While previous work showed that LLM-Chains offer some control to non-AI-experts~\cite{prompt-chainer}, our findings revealed their shortcomings when designing chatbots for cyberbullying education. Overcoming these limitations will require addressing them from multiple directions.

On the individual response level, such as when dealing with a specific chatbot reply, there are existing LLM techniques that can aid in controlling the generated text. One such technique involves adjusting the "temperature" parameter of an LLM, which serves as a rudimentary yet established means to regulate the variability of the generated text. A higher temperature value results in more "creative" output. One can also restrict text generation to predefined user concepts~\cite{stowe22ControlledGeneration}. This could ensure that the chatbot improvises freely while remaining within a positive context, like P6 requested. Incorporating control codes can further facilitate the enforcement of specific text generation patterns~\cite{keskar19ControlCodes}. While these approaches have been evaluated from an NLP perspective, future work must address their integration into the chatbot design process.

When it comes to shaping the flow of a conversation, various approaches are available. Prior research has indicated that prompting can guide a conversation to some extent, but it remains challenging to provide precise guidance, especially for non-AI-experts~\cite{herdingAIcats,zamfirescu2023johnny}. Our findings showed that LLM-Chains with predefined dialogue flows grant teachers more detailed control, albeit limiting the guidance on a more abstract level. For instance, our participants could not specify that a chatbot should dwell on a topic for a certain duration before transitioning to a new subject, all while considering the student's behavior. It is an open question how a system should be designed to enable teachers to steer the chatbot while preserving its capacity for improvisation within predefined guidelines.

The concept of guided improvisation also raises the broader question of how much control teachers are willing to relinquish in favor of encouraging improvisation. Our study demonstrated that current tools empower teachers to construct chatbots that can improvise, and teachers expressed a desire for variability in the chatbot's responses to catalyze educational outcomes. However, it is essential to recognize that granting the chatbot more flexibility increases the risk of unintended behavior. This issue is particularly relevant when teaching sensitive subjects like bystander interventions to cyberbullying. Further research is necessary to understand where teachers should draw the line between improvisation and control.

Besides the additional "levers" needed in LLMs to achieve more controlled improvisation,  additional pedagogical solutions should be considered to address LLMs' limitations and ensure students' emotional well-being while handling sensitive topics like cyberbullying: 1) Scaffolding: guiding students on how to interact with the chatbot, offering hints or prompts when needed, and providing frameworks or structures to prevent the conversations from going awry. 2) Monitoring: observing how students engage with the chatbot, making sure the language being used is age-appropriate and aligns with  teens’ emotional and cognitive development stages. 3) Debriefing: conducting debriefing sessions to help students process what they have learned, discuss their experiences, and address the emotional and psychological impacts of the chatbot intervention. 

\subsection{Multi-Participant Role-Play with Chatbots}
While the concept of multi-player improvisation theatre has been explored in role-playing games~\cite{jones2021chasing}, the guided improvisation could open up room for multi-participant role-play where one or multiple students could interact with a single or multiple chatbots playing different roles. This kind of rich environment with multi-participant interactions and interpretations would resemble interactions on social media platforms where cyberbullying exchanges are played out in front of other users who can attenuate (e.g., by supporting a victim) or amplify (e.g., by staying silent or resharing an offensive message) the effects of cyberbullying through their actions~\cite{difranzo2018upstanding}. Blending real participants and imagined identities enacted by chatbots could help youth practice socio-emotional skills in various relational and situational contexts, e.g., involving social circles of friends and peers, being part of a group or a sole upstander, interacting with people of similar or diverse views and identities, etc. As mentioned earlier, bystanders' sense-making, reading of contextual cues, emotional reactions, and anticipated consequences of their actions are tethered to social and peer contexts in which they reside~\cite{casey2017situational}, and multi-participant interactions could provide opportunities for collaborative role-playing practices and learning.

From a technical standpoint, LLMs have been used to stage social simulacra~\cite{park22SocialSimulacra,park23Simulacra}. These social interactions of multiple participants are reminiscent of the role-play scenarios our teachers envisioned. LLM-based social simulacra could, therefore, be an opportunity for bringing teachers' role-play ideas to life. It is, however, still an open question how teachers can keep control of the simulations and how the students can interact with the simulated roles. 

From the instructional perspective, chatbot role-playing sessions with multiple student participants would need to be carefully implemented and build on skills previously practiced in single-user chatbot interactions. In other words, the teachers would have to assess whether and when students are ready to move from single-user to multi-user interactions. Furthermore, because of greater autonomy and improvisation afforded in multi-participant interactions, teachers would need to be more closely involved through monitoring, moderation, and debriefing of these exchanges. Thus, there is a trade-off between improvisation and control, and greater improvisation in chatbot interactions would have to be counterbalanced by teachers' involvement in other ways. 

\section{Conclusion}

In this work, we explore what technical and design components teachers need to build chatbots that assist in bystander education through \system, an LLM-Chain based, no-code chatbot design tool. To create chatbot tools that fulfill teachers' needs, tool designers will want to consider the teachers' goal of constructing role-play scenarios and their perception of being playwrights of these social interactions. Teachers want to control and adapt the chatbot while at the same time allowing the chatbot enough improvisation so that students can explore different bystander actions and scenarios and practice socio-emotional skills. This view helps to understand how far current language model technology can be utilized for chatbot building and what new solutions still need to be found. We hope that researchers and designers of future tools will consider these factors to ensure that chatbots for adolescent cyberbullying education have a successful impact in the classroom.

\begin{acks}
The authors would like to thank all participating teachers for their valuable time and insights, the research assistants Ashley Yu, George Gu, Jade Yang, Jerry Guo, Kyle Lou, Morgan Cupp, and Tony Yang for their help in developing the probe, as well as Dominic DiFranzo and Winice Hui for their contributions to the study. This work is supported by National Science Foundation under grants IIS-2313077 and IIS-2302977. Qian Yang is also supported by Schmidt Futures’ AI2050 Early Career Fellowship.
\end{acks}

%% file: paper_appendix.tex
\section{Student Simulations}

The student simulations used \texttt{GPT-3.5-Turbo} to generate student answers when testing the chatbot. Prompts were used that contained the context along with a description of the student type. We defined three student types: an aggressive student bullying the bully, an upstander student supporting the victim, and a passive bystander student that ignores the bullying. All prompts had the same structure and only replaced parts that related to the student type. A separate prompt was used to generate the comment and any answers to the chatbot.

\subsection{Aggressive student}
\begin{verbatim}
You are John, an aggressive student, and you see the
following on Instagram: {general_context}

----

Give a comment that the student John would post under the 
Instagram post in which John insults the bully. Be 
aggressive. Answer in the language style of a teenager.
Give an answer that is no longer than 10 words.

----

You commented under this Instagram post the following
comment {comment}. 
Based on your comment, a chatbot is trying to teach you
how to best act with a cyberbullying situation. This is
your conversation so far: {messages}.

Give the next answer of the student to this conversation
where you tend to not agree with the chatbot.
Answer in the language style of a teenager. Give an answer
that is no longer than 10 words.\nJohn: 
\end{verbatim}

\subsection{Upstanding student}
\begin{verbatim}
You are John, a supportive student, and you see the 
following on Instagram: {general_context}

----

Give a comment that the student John would post under the 
Instagram post in which John comforts and supports Alex 
(the victim). Be gentle and sweet. Answer in the language 
style of  a teenager. Give an answer that is no longer 
than 10 words.

----

You commented under this Instagram post the following
comment {comment}. 
Based on your comment, a chatbot is trying to teach you
how to best act with a cyberbullying situation. This is
your conversation so far: {messages}.

Give the next answer of the student to this conversation
where you tend to agree with the chatbot.
Answer in the language style of a teenager. Give an answer
that is no longer than 10 words.\nJohn: 
\end{verbatim}

\subsection{Passive bystander student}
\begin{verbatim}
You are John, a student who ignores the bullying and
just comments on the original post, and you see the 
following on Instagram: {general_context}

----

Give a comment that the student John would post under 
the Instagram post in which John is looking forward 
to seeing the ballet recital. Be gentle and sweet. 
Answer in the language style of a teenager. Give an 
answer that is no longer than 10 words.

----

You commented under this Instagram post the following 
comment {comment}. Based on your comment, a chatbot is
trying to teach you how to best act with a cyberbullying 
situation. This is your conversation so far: {messages}.

Give the next answer of the student to this conversation
where you tend to agree with the chatbot.
Answer in the language style of a teenager. Give an 
answer that is no longer than 10 words.\nJohn: 
\end{verbatim}

\section{Student Behavior Components}

The structure and examples provided by the teacher were used to build few-shot classifiers of the student behavior. All behavior components that had the same parent component were used as classes in a classifier. The following prompt was used with a loop over all examples:

\begin{verbatim}
Victim's name is Alex. Bully's name is Leslie. 
Classify the user inputs into one of the following 
categories: 
{prompt_classes}

Only give the name of the category. If none of these 
categories match, output 'none' as category'.

Input {example_num}: {example}
Category {example_num}: {class_name}

Input {example_num}: {student_message_to_classify}
Category {example_num}: 
\end{verbatim}

We used \texttt{Text-Davinci-003} and parsed its answer to determine the predicted class (and therefore the conversational path to take in the dialogue structure).

\section{Chatbot Reaction Component}
The response examples provided by the teacher were used to generate the chatbot's answer in each situation. As example contexts, the behavior examples from the parent student behavior component were used. We prompted \texttt{Text-Davinci-003} for the generation with a loop over all teacher-defined examples.

\begin{verbatim}
The student sees a cyberbully on social media. 
The bully's name is Leslie and the victim's name is 
Alex. 
The student makes a comment in response to the post. 
You are talking to that student whose name is not Alex
or Leslie so don't call him/her Alex or Leslie. 
Teach that student to counteract cyberbullies based on
the following examples:"

Example: {example_num}
Context: {context_example}
Response: {response}"

Now fill in a new response based on the examples.
Give answers very similar to the examples:

Context: {student_message_to_answer}
Response: 
\end{verbatim}